# Understanding the neural architecture of emotion regulation by comparing two different strategies: A meta-analytic approach

Bianca Monachesi[1,], Alessandro Grecucci[1,2], Parisa Ahmadi Ghomroudi[1], Irene Messina[3]

[1] *Clinical and Affective Neuroscience Lab, Department of Psychology and Cognitive Sciences - DiPSCo, University of Trento, Rovereto, Italy*
[2] *Center for Medical Sciences - CISMed, University of Trento, Trento, Italy*
[3] *Universitas Mercatorum, Rome, Italy*

**Abstract**

In the emotion regulation literature, the amount of neuroimaging studies on cognitive reappraisal led the impression that the same top-down, control-related neural mechanisms characterize all emotion regulation strategies. However, top-down processes may coexist with more bottom-up and emotion-focused processes that partially bypass the recruitment of executive functions. A case in point is acceptance-based strategies. To better understand neural commonalities and differences behind different emotion regulation strategies, in the present study we applied a meta-analytic method to fMRI studies of task-related activity of reappraisal and acceptance. Results showed increased activity in left-inferior frontal gyrus and insula for both strategies, and decreased activity in the basal ganglia for reappraisal, and decreased activity in limbic regions for acceptance. These findings are discussed in the context of a model of common and specific neural mechanisms of emotion regulation that support and expand the previous dual-routes models. We suggest that emotion regulation may rely on a core inhibitory circuit, and on strategy-specific top-down and bottom-up processes distinct for different strategies.

**Introduction**

In affective neuroscience and in clinical psychology, emotion regulation (Gross, 1998) has emerged as a core construct widely applied to the conceptualization of neurobiological models of affective disorders (Kring & Sloan, 2009; Taylor & Liberzon, 2007; Grecucci et al., 2020; Messina et al., 2021) and their treatment (Beauregard, 2007; Messina et al., 2013; Frederickson et al., 2018; Grecucci et al., 2015; 2017). In parallel with this growing scientific interest on emotion regulation, the research has seen a rising debate regarding the usefulness of different emotion regulation strategies and their implications for therapeutic techniques (Dadomo et al., 2016; 2018; Grecucci et al., 2017; Leahy et al., 2011; Wolgast et al., 2011). In this debate, reappraisal and acceptance are often mentioned as effective strategies to regulate emotions and mechanisms of psychotherapy action (Wolgast et al., 2011; Grecucci et al., 2020; Spencer et al., 2020).

Reappraisal is defined as "*construing a potentially emotion-eliciting situation in non-emotional terms*" (Gross, 2002, p.281). It has been traditionally deemed adaptive since associated with reduced neuropsychological response to emotional events (e.g., Webb et al., 2012; Kanske et al. 2012), and with general well-being and mental health (Aldao et al. 2010). It allows people to change the appraisals that contribute to negative emotions (Gross, 1998), by recruiting a certain amount of cognitive resources as reflected in the involvement of a complex pattern of prefrontal cortical regions (Ochsner & Gross, 2005). Reappraisal is clearly related to traditional cognitive behavioural therapy (CBT), which use cognitive restructuring to alleviate psychological suffering through changing how the patient interpret and think his/her experiences in everyday life (Beck et al. 1979). We acknowledge that different types of reappraisal strategy exist (i.e., reinterpretation and distancing), and that previous studies highlighted that they rely on partial distinct mechanisms as well as cortical brain areas (Messina et al., 2015; Power & LaBar, 2019). Here, then, we will focus only on the reinterpretation strategy - referred to as reappraisal hereafter, and intended as the reappraised situation or the cause of the stimulus, without any change in the perspective taken.

On the other hand, acceptance can be described as a mental stance toward ongoing mental and sensory experience, characterized by openness, curiosity, and non-evaluative attitude (Grecucci et al., 2015; Goldin et al., 2019) as well as by the recruitment of very few cognitive resources relying on prefrontal cortical areas (Messina et al., 2021). Acceptance is the core of the so called third wave behavioural therapies (Kahl et al., 2011; Hayes, 2004). In this context, it has been described as "*the active and ware embrace of private experiences without unnecessary attempts to change their frequency or form*" (Hayes et al., 2012, p.982) and it is taught as the counter of experiential avoidance. More implicitly, also psychodynamic and humanistic approaches work on experiential avoidance/acceptance encouraging the experience of emotions and the associated impulse physically in the body, rather than down-regulating them through cognitive or attentional mechanisms (Frederickson et al., 2018; Messina et al., 2020; 2021).

In terms of psychophysiological effects, both reappraisal and acceptance are widely considered adaptive strategies (Aldao et al., 2010). Previous studies which have experimentally compared these strategies have reported that they are effective in reducing experimentally inducted negative emotions and physiological activation, but at the same slight differences have emerged. When comparing their efficacy in reducing short-term negative emotions, in most cases reappraisal was slightly superior to acceptance (Goldin et al., 2019; Hofmann et al., 2009; Smoski et al., 2015; Szasz et al., 2011; Troy et al., 2018), even if other studies found no significant differences (Asnaani et al., 2013; Wolgast et al., 2011). With regard to physiological reactivity, Hofmann and colleagues (2009) have reported similar effectiveness of both strategies in decreasing heart rate (when compared to suppression). Goldin and colleagues (2019) reported no difference in respiration rate and skin conductance, but higher heart rate in reappraisal compared to acceptance. Wolgast and colleagues (2011) found that reappraisal was slightly more effective than acceptance at reducing skin conductance, whereas Troy and colleagues (2018) reported the opposite result. Finally, only one study (Troy et al., 2018) have examined the perceived cognitive costs of using these two strategies reporting that acceptance was perceived as less difficult to deploy than reappraisal.

Although these results suggest that both reappraisal and acceptance can be considered useful strategies, the underlining neurobiological mechanisms are still poorly understood. The investigation of common and different brain regions associated with reappraisal and acceptance not only may clarify their specific nature, but it may crucially unveil those control-related brain areas underpinning top-down vs bottom-up (emotion focused) strategies, contributing then to a deeper understanding of the mechanisms of emotion regulation. Indeed, traditional models of emotion regulation are largely based on, and overlap with the neural structures involved in reappraisal (Ochsner & Gross, 2005), despite the growing body of evidence of more emotion-focused regulation modalities (Messina et al., 2021). A recent study (Messina et al., 2021) has pointed out that its neural correlates may differ from reappraisal, with a much less clear relevance of prefrontal control brain areas, and possibly recruiting more bottom-up mechanisms. Unfortunately, this study didn't report a comparison between acceptance and reappraisal, so the possible differences between the two strategies remain speculative.

To fit this emergent literature, some authors have proposed a dual-route model for emotion regulation for which different top-down cognitive control mechanisms and bottom-up emotion focused mechanisms are possible (e.g., Grecucci et al., 2020). However, dual-route models may be simplistic, and one intriguing hypothesis is that there might also be a common mechanism behind different strategies (Morawetz et al., 2017). To date, only four task-based fMRI experiments have directly compared reappraisal and acceptance (Goldin et al., 2019; Dixon et al., 2020; Opialla et al., 2015; Smoski et al., 2015). In most of these studies, greater brain responses in prefrontal brain regions implicated in cognitive control (i.e., dorso-lateral prefrontal cortex, DLPFC, and dorso-medial prefrontal cortex, DMPFC) have been observed in reappraisal compared to acceptance (Goldin et al., 2019; Dixon et al., 2020; Smoski et al., 2015). In some of these studies, acceptance has been associated with reduced activity in parts of the default mode network (DMN) (Dixon et al., 2020; Opialla et al., 2015). The DMN is a set of areas anti-correlated to executive processes and associated with mind-wandering (Christoff et al., 2009). Since mind-wandering has been considered as the opposite of mindfulness (Mrazek et al., 2012), these effects on DMN have been interpreted as due to the interruption of ruminative, self-reflective processes over emotions, which are independent from executive processes (Ellard et al., 2021; Messina et al., 2021). Finally, Smoski and colleagues (2015) also reported greater activation in regions linked to somatic and emotion awareness (left insular cortex and right prefrontal gyrus) in acceptance compared to reappraisal. In other words, these studies suggest that reappraisal and acceptance may rely on different neural substrates: on one hand, a regulatory mechanism based on cognitive control (reappraisal) and supported by prefrontal executive regions. On the other hand, an acceptance-based mechanism which acts without the involvement of executive areas and based on the reduction of brain activity in subcortical areas and of the DMN. However, a few experiments reported increased prefrontal activations for acceptance (Lebois et al., 2015; Goldin et al., 2019). It follows that a common core mechanism may exist independently from the strategy used.

To provide evidence on this issue, in the present meta-analytic study, we made an initial attempt to compare fMRI studies of reappraisal and acceptance in order to shed light on the possible common and distinct neural mechanisms underlying them. Consequently, these results may also shed light on the potential mechanisms involved in these two types of strategies. Reappraisal-based strategies have been always considered to rely on control-related (or "top-down") regulation mechanisms, whereas acceptance-based strategies have been conceptualized as relying on emotion focused (or "bottom-up") regulation mechanisms (Grecucci et al., 2020; Messina et al., 2022). Showing that these two strategies rely on different neural mechanisms may indicate that they rely on different psychological mechanisms too.

In the present study, we aim to explore the hypothesis that strategy specific mechanisms may co-exist with a common core mechanism. To this aim, we used a coordinate-based Activation Likelihood Estimation (ALE) method (Laird et al., 2005) to quantitatively compare two sets of functional Magnetic Resonance Imaging (fMRI) studies which have contrasted whole-brain activity in reappraisal and/or acceptance conditions relative to baseline (control condition, in which no regulation was performed). Namely, we performed a conjunction analysis to observe possible core common regulation mechanism involved in both strategies, along with a contrast analysis to examine the existence of significant clusters of brain activity that are specific of each of the two strategies. For both conjunction and contrast analysis, regions of increased and decreased activity were explored.

Although many previous meta-analytic studies on emotion regulation strategies have been conducted, we believe our approach has the potential to better understand commonalities and differences in emotion regulation strategies. On the basis of the existing literature and a previous metanalytic study (e.g., Morawetz et al., 2017), in which many types of strategies were pooled, we hypothesize that the ventro-lateral prefrontal cortex (VLPFC) and insula may be good candidates for a core common mechanism, for their strategic position in inhibiting emotion related areas, and for the implication in language function (semantic and phonological processes), especially the left portion. In addition to this, we believe, reappraisal-based strategies may recruit large dorso-lateral portions of the prefrontal cortex (Buhle et al., 2014), and acceptance-based strategies, more subcortical limbic structures (Messina et al., 2021).

**Methods**

*Study selection*

The authors selected studies through a systematic online search on PubMed (https://www.ncbi.nlm.nih.gov/pubmed), up to the August 2022. The keywords used for the online search were 'emotion regulation,' 'emotion regulation strategies' AND 'reappraisal,' 'acceptance,' and/or 'mindfulness' AND 'fMRI' or 'neuroimaging.' References of the retrieved studies, as well as relevant previous reviews or systematic reviews, and meta-analyses, were hand-searched for further supplement. The entire literature screening process followed the guidelines of PRISMA (Page et al., 2021, see Supplementary material S1 for the PRISMA 2020 checklist and Figure 1 for the PRISMA flowchart). No previous registration or protocol was prepared.

The initial selection focused on studies that employed the typical emotion regulation task, where a condition of emotion regulation is compared to a control condition of no-regulation during the presentation of emotional stimuli. The following inclusion criteria were applied:

- studies which reported the specific contrasts of emotion regulation (acceptance/reappraisal) > no-regulation and/or the no-regulation > emotion regulation (acceptance/reappraisal);
- studies in which only univariate whole-brain analysis was performed (studies or analysis using ROI approach were excluded to avoid inflated results, Muller et al., 2018);
- studies in which Montreal Neurological Institute (MNI) and Talairach coordinates were reported;
- studies performed only on adults (18-55 years old), drug-free participants, and with no neurological diseases.

Exclusion criteria from the retrieved studies were (i) unclear or not specific reinterpretation strategy (i.e., reappraise the situation or the cause of the stimulus, without any change in perspective taking, as in distancing strategy), (ii) no separate contrast for negative stimuli, (iii) no separate contrast for down-regulation, (iv) no significant foci (see Muller et al., 2018 for the sensitivity of coordinate-based algorithm to non-significant results), and (v) no provided information when requested. For an overview of the specific instructions used in the acceptance studies, see Messina et al., (2021). The final dataset included 32 studies among those investigating acceptance and/or reappraisal. Studies with more than one relevant contrast, or with separate analysis between conditions/participants' groups were considered as independent samples for a total of 50 records included in the meta-analysis. These are all reported in Table 1 and marked as "a" and "b" (and so on) when belonging to the same study.

**Table 1. Studies included in the meta-analysis. F = female; acc = acceptance strategy; reap = reappraisal strategy; M = Mean; SD = Standard Deviation. Articles with more than one relevant contrast is reported as "a", "b" and so on. (*) means data provided by authors on request.**

| Studies | Contrast | N | Age M(SD) | N foci |
| --- | --- | --- | --- | --- |

| | | | | | |
|---|---|---|---|---|---|
| 1 | Lutz et al., 2014a | acc vs no-regulation | 24 | 29.98(7.96) | 3 |
| 2 | Smoski et al., 2015a | acc vs no-regulation | 19 (12 F) | 27.9 (6.3) | 8 |
| 3 | Smoski et al., 2015b | acc vs no-regulation | 19 (12 F) | 27.9 (6.3) | 5 |
| 4 | Murakami et al., 2015 | acc vs no-regulation | 21 (11 F) | 25.1 (5.5) | 22 |
| 5 | Lebois et al., 2016a | acc vs no-regulation | 30 (15 F) | 18–23 | 8 |
| 6 | Ellard et al., 2017a | acc vs no-regulation | 21 F | 29.48 (8.44) | 2 |
| 7 | Goldin et al., 2019a | acc vs no-regulation | 35 (57% F) | 32.2 (8.9) | 11 |
| 8 | Dixon et al., 2020* | acc vs no-regulation | 113 (61 F) | 32.9 (7.92) | 2 |
| 9 | Kross et al., 2009a | no-regulation vs acc | 24 (15 F) | 20.83(3.27) | 64 |
| 10 | Lutz et al., 2014b | no-regulation vs acc | 24 | 29.98(7.96) | 2 |
| 11 | Lebois et al., 2016b | no-regulation vs acc | 30 (15 F) | 18–23 | 1 |
| 12 | Ellard et al., 2017b | no-regulation vs acc | 21 F | 29.48 (8.44) | 10 |
| 13 | Kober et al., 2019a | no-regulation vs acc | 17 (5 F) | 31.75(5.18) | 3 |
| 14 | Kober et al., 2019b | no-regulation vs acc | 17 (5 F) | 31.75(5.18) | 9 |
| 15 | Goldin et al., 2019b | no-regulation vs acc | 35 (57% F) | 32.2 (8.9) | 4 |
| 16 | Dixon et al., 2020b* | no-regulation vs acc | 113 (61 F) | 32.9 (7.92) | 9 |
| 17 | Dixon et al., 2020c* | no-regulation vs acc | 35 (22 F) | 32.1(8.70) | 6 |
| 18 | Westbrook et al., 2013 | no-regulation vs acc | 48 (31% F) | 45 (11.35) | 1 |
| 19 | Che et al., 2015 | reap vs no-regulation | 29 (15 F) | 22.62 (1.59) | 8 |
| 20 | Dixon et al., 2020d* | reap vs no-regulation | 35 (22 F) | 32.1(8.70) | 19 |
| 21 | Dorfel et al., 2014 | reap vs no-regulation | 19 F | 18 -39 | 17 |
| 22 | Fitzgerald et al., 2020 | reap vs no-regulation | 49 (67% F) | 25.24(7.98) | 13 |
| 23 | Gianaros et al., 2014a | reap vs no-regulation | 157 (88 F) | 30–54 | 21 |
| 24 | Goldin et al., 2008 | reap vs no-regulation | 17 F | 22.7 (3.5) | 18 |
| 25 | Goldin et al., 2019c | reap vs no-regulation | 35 (57% F) | 32.2 (8.9) | 13 |
| 26 | Golkar et al., 2012 | reap vs no-regulation | 58 (32 F) | 24.02 (2.26) | 11 |
| 27 | Harenski et al., 2006 | reap vs no-regulation | 10 F | 18–29 | 7 |
| 28 | Herwig et al., 2007a | reap vs no-regulation | 18 | 23–36 | 2 |
| 29 | Macdonald et al., 2020a | reap vs no-regulation | 19 | 27 | 8 |
| 30 | Morawetz et al., 2016 | reap vs no-regulation | 59 (20 F) | 32.47(11.25) | 2 |
| 31 | New et al., 2009 | reap vs no-regulation | 14 F | 31.7 (10.3) | 14 |
| 32 | Ochsner et al., 2002 | reap vs no-regulation | 15 F | 21.9 | 12 |
| 33 | Qu et al., 2017a | reap vs no-regulation | 29 (14 F) | 19.2 | 11 |
| 34 | Silver et al., 2015 | reap vs no-regulation | 30 (13 F) | 21.97 | 48 |
| 35 | Simsek et al., 2017 | reap vs no-regulation | 15 F | 22.53 (1.80) | 8 |
| 36 | van der Velde et al., 2015 | reap vs no-regulation | 51 (47 F) | 37.1 (10.3) | 21 |
| 37 | vanderhasselt et al., 2013 | reap vs no-regulation | 42 F | 21.26 (2.29) | 7 |
| 38 | Wager et al., 2008 | reap vs no-regulation | 30 (18 F) | 22.3 | 8 |
| 39 | Wu et al., 2019 | reap vs no-regulation | 15 | 21 - 27 | 10 |
| 40 | Yoshimura et al., 2014a | reap vs no-regulation | 15 (9 F) | 23.3(2.2) | 7 |
| 41 | Ziv et al., 2013a | reap vs no-regulation | 27 (12 F) | 31.1 (7.6) | 11 |
| 42 | Ziv et al., 2013b | reap vs no-regulation | 27 (12 F) | 31.1 (7.6) | 1 |
| 43 | Gianaros et al., 2014b | no-regulation vs reap | 157 (88 F) | 30–54 | 17 |
| 44 | Goldin et al., 2019d | no-regulation vs reap | 35 (57% F) | 32.2 (8.9) | 1 |
| 45 | Herwig et al., 2007b | no-regulation vs reap | 18 | 23–36 | 3 |

| 46 | Koenigsberg et al., 2010 | no-regulation vs reap | 16 (9 F) | 31.8(7.7) | 5 |
| 47 | Kross et al., 2009b | no-regulation vs reap | 24 (15 F) | 20.83(3.27) | 5 |
| 48 | Macdonald et al., 2020b | no-regulation vs reap | 19 | 27 | 5 |
| 49 | Qu et al., 2017b | no-regulation vs reap | 29 (14 F) | 19.2 | 2 |
| 50 | Yoshimura et al., 2014b | no-regulation vs reap | 15 (9 F) | 23.3(2.2) | 3 |

*ALE Analyses procedure*

The Activation Likelihood Estimation (ALE) method (Eickhoff et al., 2009) is based on an algorithm able to overcome spatial uncertainty associated with neuroimaging studies, treating each focus coordinates as the center of a Gaussian spatial probability distribution. The resulting ALE maps consist in the spatial convergence of activation probabilities across experiments foci. Permutation procedure allows noise (random clustering) to be distinguished from the true convergence of foci. All analyses in the present study were carried out using the GingerALE v3.02 software (http://brainmap.org/). Before performing the conjunction and contrast analyses, all foci were converted in MNI coordinates using icbm2tal transform (Lancaster et al., 2007). We then run separate ALE analyses on the considered subsets: (a) reappraisal versus no-regulation, to obtain the ALE map of increased brain activity in reappraisal; (b) no-regulation versus reappraisal, to obtain the ALE map of decreased brain activity in reappraisal; (c) acceptance versus no-regulation, to obtain the ALE map of increased brain activity in acceptance; (d) no-regulation versus acceptance, to obtain the ALE map of decreased brain activity in acceptance. For each separate analysis, the statistical significance was assessed and corrected for multiple comparisons by a cluster-level family-wise error (FWE) method, which represents the best compromise between sensitivity and specificity (Eickhoff et al., 2016; Müller et al., 2018). An uncorrected cluster-forming threshold of $p = .01$, a cluster-level inference threshold (controlling the reliability of cluster size, Eickhoff et al., 2012) of $p = 0.05$, and 1000 permutations were used.

Once we obtained the four ALE images, the contrast and the conjunction analysis were computed between both the two strategies (acceptance/reappraisal) > no-regulation subsets (map of increased activity), and the two no-regulation > strategies (acceptance/reappraisal) subsets (map of decreased activity). The contrast analysis creates two ALE contrast images by directly subtracting one ALE image from the other. The conjunction analysis shows the similarity between the datasets using the voxel-wise minimum value of the ALE images. Study size correction (Eickhoff et al., 2012) is performed by GingerALE pooling all foci datasets and randomly dividing them into two groups of the same size as the original data sets. An ALE image is created for each new data set, then subtracted from the other and compared to the true data. After 1000 permutations, a voxel-wise P value image showed the location of the true data's values on the distribution of values in that voxel. The results were thresholded with $p = .01$. A default cluster size $> 200$ mm³ was applied. Cluster analysis of contrast images uses Z score values. Surf Ice software was used to plot the resulting brain maps (https://www.nitrc.org/projects/surfice/).

## Results

*Included studies and samples characteristics*

The dataset of acceptance included 8 studies (for a total of 281 participants) reporting results for the contrast acceptance vs. no-regulation, yielding a total of 61 foci of *increased* brain activity in acceptance. Ten studies (for a total of 364 participants) reported the contrast no-regulation vs. acceptance, yielding a total 109 foci of *decreased* brain activity in acceptance. The ALE analysis of reappraisal, instead, included 24 studies (for a total of 815 participants) reporting the contrast reappraisal vs. no-regulation, yielding a total of 297 foci of *increased* brain activity in reappraisal, and 8 studies (for a total of 305 participants) reporting the contrast no-regulation vs. reappraisal, yielding a total of 41 foci of *decreased* brain activity in reappraisal.

For reasons of completeness, the resulting ALE maps for each single meta-analysis are presented in the Supplementary material Table S2 and Table S3 for acceptance and reappraisal results, respectively. When uncorrected cluster-forming threshold of $p = 0.01$ did not yield significant foci (i.e., for acceptance and no-regulation vs. reappraisal results), we applied a more lenient threshold of $p = 0.05$ to allow GingerAle to perform the conjunction and contrast analysis.

*Common neural mechanisms for reappraisal and acceptance (Conjunction Analysis)*

The conjunction analysis of common increased brain activity during reappraisal and acceptance revealed three clusters of significant brain activation. Two clusters were located in the inferior frontal gyrus (or VLPFC), whereas one cluster was in VLPFC and insula (see Figure 2). No shared clusters of decreased brain activity emerged between reappraisal and acceptance (also when results were thresholded with more lenient *p* = .05).

**Table 2. Common neural mechanisms for reappraisal and acceptance. Coordinates x, y, z of local maxima refer to MNI-space. BA = Brodmann Area; L= left.**

| Cluster | x | y | z | ALE | Label | Cluster size (mm$^3$) |
|---|---|---|---|---|---|---|
| 1 | -38 | 24 | -6 | 0.009 | L insula | 984 |
|   | -50 | 18 | -6 | 0.003 | L inferior frontal gyrus | |
| 2 | -50 | 20 | 12 | 0.002 | L inferior frontal gyrus (BA 45) | 24 |
| 3 | -52 | 22 | -14 | 0.009 | L inferior frontal gyrus (BA 45) | 24 |

**Specific neural mechanisms for reappraisal and acceptance** *(Contrast Analyses)*

When the ALE maps of reappraisal and acceptance were contrasted, two different clusters of increased and two different clusters of decreased (results thresholded with more lenient *p* = .05) brain activity emerged for reappraisal vs acceptance. Increased activity was located in the superior frontal gyrus (cluster 1) and in the left middle frontal gyrus (cluster 2), whereas decrease brain activity involved left globus pallidus (cluster 1) and left putamen (cluster 2) (see Table 3, Figure 3).

**Table 3. Specific neural mechanisms for reappraisal. Coordinates x, y, z of local maxima refer to MNI-space. BA = Brodmann Area; L= left.**

| Cluster | x | y | z | P | Label | Cluster size (mm$^3$) |
|---|---|---|---|---|---|---|
| **a. Increased brain activity** | | | | | | |
| 1 | -14 | 22.3 | 56.3 | 0 | L superior frontal gyrus (BA 6) | 2008 |
|   | -8.6 | 20.4 | 59.2 | 0.001 | L superior frontal gyrus (BA 6) | |
|   | -.5 | 21 | 62 | 0.002 | L superior frontal gyrus (BA 6) | |
| 2 | -34 | 9.5 | 44 | 0.007 | L middle frontal gyrus (BA 6) | 304 |
| **b. Decrease brain activity** | | | | | | |
| 1 | -16 | 1 | -15 | 0.023 | L globus pallidus | 1032 |
| | -20.5 | -2.7 | -12.1 | 0.046 | L globus pallidus | |
| 2 | -32 | -10 | -6 | 0.023 | L putamen | 240 |
| | -27 | -9.3 | -9.6 | 0.046 | L putamen | |

In addition, one cluster of increased and two clusters of decrease (results thresholded with more lenient $p = .05$) brain activity emerged as specific for acceptance vs reappraisal. The former was located in the claustrum, whereas the latter involved bilaterally the posterior cingulate (cluster 1), the right parahippocampal gyrus, and the right thalamus (cluster 2) (see Table 4, Figure 3).

**Table 4. Specific neural mechanisms for acceptance Increased (a) and decrease (b) brain activity in the contrast analysis between acceptance vs reappraisal. Coordinates x. y. z of local maxima refer to MNI-space. BA = Brodmann Area; L= left; R=right.**

| Cluster | x | y | z | P | Label | Cluster size (mm³) |
|---|---|---|---|---|---|---|
| **a. Increased brain activity** | | | | | | |
| 1 | -29 | 18 | 7 | 0.016 | claustrum | 424 |
| | -32 | 14 | 7.2 | 0.006 | claustrum | |
| **b. Decrease brain activity** | | | | | | |
| 1 | 11 | -50 | 14 | 0.009 | R posterior cingulate (BA 30) | 1528 |
| | 12.9 | -53 | 18.1 | 0.024 | R posterior cingulate (BA 30) | |
| | -3.8 | -52.8 | 9.1 | 0.031 | L posterior cingulate (BA 30) | |
| 2 | 16.4 | -39.3 | 6.9 | 0.024 | R parahippocampal gyrus (BA 30) | 232 |
| | 18 | -34 | 6 | 0.04 | R pulvinar | |
| | 11 | -40 | 5 | 0.045 | R parahippocampal gyrus (BA 30) | |

**Discussion**

Despite thirty years of research in emotion regulation, a clear understanding of the neural basis of emotion regulatory processes has not emerged yet. This is because, during the years, scientists have focused only on a subset of strategies, e.g. reappraisal-based strategies, thus giving the impression that one unique neural substrate characterize all emotion regulation strategies. However, in the last years, scholars started studying a different and quite opposite set of strategies to regulate emotions, e.g. acceptance-based strategies (Campbell-Sills et al., 2006; Hofmann et al., 2009; Wolgast et al., 2011; Grecucci et al., 2015; Messina et al., 2021), also named emotion-focused or experiential strategies in other contexts (Greenberg & Vandekerckhove, 2009; Grecucci et al., 2020). Importantly, these classes of strategies are believed to rely upon different psychological (Messina et al., 2021). As such, a comparison between the neural bases of these two types of strategies may indicate that they rely on different psychological mechanisms too. Riding the wave of such evidence, an intriguing hypothesis is that emotion regulation processes rely on strategy-specific mechanisms which play in concert with partially overlapping mechanisms (a core regulatory process).

To shed light on these points, in the present paper we meta-analytically compared the neural bases of reappraisal and acceptance strategies and we found preliminary evidence in support of common and – most importantly, of separate neural substrates (Grecucci et al., 2020). In the following sections we discuss these findings by starting from the common core mechanism, and then outlining strategy-specific mechanisms.

*Common regulatory processes*

When performing the conjunction analyses, we confirmed that both acceptance and reappraisal showed increased activations in common areas, the VLPFC and in the insula. The VLPFC is implicated in response interpretation, selection and action inhibition as well as in semantic and phonological processing (Morawetz et al., 2016b). The insula, instead, plays a critical role in integrating sensory input from the internal and external world, in order to shape a consistent and aware representation of the inner emotional state (e.g., Zaki et al., 2012) and to map arousing states associated with emotions (Grecucci et al., 2013a,b). Indeed, a study from Grecucci and collaborators (2013a,b) showed that the insula was one of the main regions modulated by emotion regulation. Both insula and VLPFC are consistently activated during successful ER processes (Li et al., 2021), and this is true regardless of the type of emotion regulation strategy investigated (Morawetz et al., 2017), including acceptance (Messina et al., 2021). The involvement of the VLPFC is in line with recent models of emotion regulation that relativize the role of executive/controlled functions in emotion regulation, and at the same time foster the importance of spontaneous, semantic and non-effortful forms of regulation (Messina et al., 2016b; Viviani, 2013; Viviani, 2014). In accord with this view, indeed, the involvement of the VLPFC in absence of core executive areas (as in the case of acceptance, see paragraph below) has been documented also in other regulation processes which can be considered more implicit (or non-controlled), such as emotional labelling (Torre & Lieberman, 2018; Tupak et al., 2014) and spontaneous avoidance (Benelli et al., 2012; Viviani et al., 2010).

Besides these commonalities, however, our data preliminary showed that reappraisal and acceptance regulate emotion activating and deactivating partially different neural regions.

*Specific mechanisms for reappraisal*

The contrast analysis confirmed that reappraisal is specifically associated with increased activity in prefrontal regions, including superior frontal gyrus or DLPFC, and middle frontal gyrus or DMPFC. Both DLPFC and DMPFC belong to a well-established network of control-related prefrontal regions previously reported in several meta-analytic studies of reappraisal (Buhle et al., 2014; Frank et al., 2014; Messina et al., 2015; Morawetz et al., 2017), and their involvement is in line with the traditional view of emotion regulation as a form of top-down, cognitive control on emotions (e.g., Ochsner and Gross 2008). In particular, the DLPFC and the DMPFC seem to contribute to emotion regulation through response inhibition and executive control (Grecucci et al, 2013a,b; Morawetz et al., 2017; Morawetz et al., 2020). Not surprisingly, the same prefrontal regions underpin other top-down strategies such as distraction (Buhle et al., 2014).

For what concerns the decreased activity related to reappraisal, we found that it specifically involved sublobar regions, that is, globus pallidus and putamen. These results are consistent with and complementary to previous meta-analytic studies (Buhle et al., 2014; Frank et al., 2014), which found same deactivations for reappraisal, and increased activations during upregulation by reappraisal. Both globus pallidus and putamen belong to the basal ganglia (BG), which have been traditionally associated with motor functions. However, basal ganglia are also part of the recognized Interoceptive Theory of Emotion (somatic marker hypothesis, Bechara, & Damasio, 2005), which posits that emotional responses are characterized by bodily component able to support the decision-making process. A recent review emphasizes the involvement of basal ganglia in affective processing via high connections with both cortical and limbic regions, which allow the organism to adapt behavioural responses to the emotional context (Pierce & Péron, 2020). Indeed, the role of BG in the reinforcement learning permits the affective value (or internal state) and behaviour to be shape and apply to successive similar emotional conditions (Pierce & Péron, 2020). Decreased activity in this area, then, can be explain as the attempt to counteract a habitual emotional response, changing the affective value previously associated with the given context, by reappraising it. Increased activity in putamen was reported in anxiety patients relative to HC (Picó-Pérez et al., 2017), and it is suggested being part of the network for cognitive action regulation (Langner et al., 2018). Similarly, connectivity analysis reported the involvement of putamen and pallidum for cognitive emotion regulation (Kohn et al., 2014).

*Specific mechanisms for acceptance*

Differently from reappraisal, and in line with our predictions, the typical network of control-related prefrontal regions was not visible for the acceptance strategy. For this later, indeed, we found specific increased brain activity only in the claustrum. This area, a thin collection of neurons placed between the insular cortex and the striatum, has been proposed as a potential area in which multimodal input are unified in a single conscious experience (Crick & Koch, 2005), due to its high connections with sensory modalities and cortical-subcortical neuromodulations. Alternatively, it might play a role in selective attention, especially in differentiating salient, relevant information from the irrelevant ones across different sensory modalities (Goll et al., 2015). Similar to the thalamus, the claustrum is suggested to focus attention, but one step later in the sensory processing, and - as a such, in a more selective way (Goll et al., 2015). The involvement of the claustrum in acceptance can be view as a form of increased awareness of bodily-sensorial states (Grecucci et al., 2015; Messina et al., 2015), and as a form of multimodal sensory filter, allowing excessive emotional reactivity to be minimized (Dixon et al., 2020; Golding et al., 2019; Wolgast et al., 2013). The limited contribution of this structure is consistent with previous studies which reported no detectable or reduced increased activity in prefrontal cortical areas in acceptance (Kross et al., 2009; Westbrook et al., 2013; Kober et al., 2019; Dixon et al., 2020; Goldin et al., 2019), and account for a view of acceptance as a form of regulation that does not imply cognitive control to alter the emotional response (Messina et al., 2021).

Results for the acceptance-related deactivations corroborate this hypothesis. Indeed, we found acceptance is associated with the reduction of brain activity in structures of the limbic lobe, namely the posterior cingulate cortex (PCC)/precuneus and the parahippocampal gyrus, and of the thalamus (pulvinar). Interestingly, these structures are different from those specifically found for reappraisal. The PCC is a key area of the default mode network (DMN), and, according to our hypothesis, the deactivation of the PCC may reflect the interruption of inner processes, including rumination and other forms of mind wondering. The functional deactivation of the PCC associated with acceptance has been previously reported in another meta-analysis of acceptance studies (Messina et al., 2021), and activation of the PCC have been reported in association to strategies somewhat opposite to acceptance (based on avoidance), such as distancing (Koenisberg et al., 2010) and distraction (Kanske et al., 2011). Interestingly, the PCC, and the DMN more in general, are engaged in semantic processing (Binder et al., 2009; Wirth et al., 2011), again supporting that also in absence of executive processes, semantic processes may serve as a form of emotion regulation.

As for PCC, also parahippocampal gyrus (PHG) has been already reported in the cluster of areas underpinning acceptance (Dixon et al., 2020; Messina et al., 2021). According to some authors (Phillips et al., 2008), it is part of a ventromedial neural system (vs the dorsal/lateral system), which is involved in the early and automatic evaluation of the emotional meaning during the regulation of the emotion-related behavioural outcome. If decreased PHG connectivity has been reported during mindfulness/meditation practice (Hernandez et al., 2018), its abnormal activity or connectivity was associated with patients with psychopathologies related to emotion dysregulation (Tak et al., 2021; Brown et al., 2020). This suggest that the reduced activity in the PHG during acceptance may reflect the reduced impact the emotional event has on the individual in terms of memory association with or trace retrieval of the stimulus (Yang et al., 2017).

*Implications and limitations*

In this study, the results support the idea that both common and distinct mechanisms exist for reappraisal and acceptance. One implication is that previous models considering one unique cognitive model behind all strategies (e.g., the Modal Model, Gross, 2005), or dual route models of emotion regulation (cognitive vs experiential) (e.g., Grecucci et al., 2020), should be integrated in a more complex model. Based on our results, we suggest that emotion regulation process relies on a common neural mechanism (possibly related to a core inhibitory function, see Figure 4, central part of the figure), which coexists with strategy-specific mechanisms separating reappraisal-like strategies (on the left), from acceptance-like strategies (on the right).

Considering emotion regulation as a set of different phenomena instead of reducing it to the cognitive control of emotions, has relevant clinical implications in terms of tailoring therapeutic interventions to specific clinical situations. For example, in presence of an overstated attempt to control mental content, stimulating an additional form of control using reappraisal-based therapeutic intervention may turn out to be detrimental (Najmi & Wegner, 2008; Purdon, 1999), whereas encouraging the adoption of a non-controlling attitude toward emotion can be more useful (Beevers et al., 1999; Marcks & Woods, 2005).

Another consideration regards the evidence that, in healthy individuals, cognitive strategies such as reappraisal are not the primary choice when emotion intensity is high (Sheppes et al., 2011), or when participants are under stress (Raio et al., 2013). Similarly, the use of reappraisal may decrease as the severity of Social Anxiety Disorder symptoms increases (Goldin et al., 2009). When the deployment of cognitive resources to regulate emotion is constrained, for instance by psychopathological status, it may be beneficial to use a different but still adaptive emotion regulation strategies. This approach should be pursued regardless of the limitations imposed by adopting a specific approach.

Beside the merits, the finding of the present study should be considered together with the limitations, especially those concerning the samples size. Due to the novelty of the field, only an exiguous number of studies have been found in the case of acceptance. For both the strategies, in addition, the exclusive selection of the studies based on the whole-brain analyses made the procedure suitable for overcoming the often pointed out limitation of inflated results due to the inclusion of ROI studies (see Müller et al., 2018; Frank et al., 2014). However, this choice had as counterpart an important reduction of the available studies (in some cases, less than the suggested 17 studies, Eickhoff et al., 2016). The exclusion of ROI-based studies may have had a further implication for the contrast no-regulation vs strategy. Relevant structures related to emotion processing, such as the amygdala, are typical regions of interest in task-related functional analyses. Many studies provide evidence that activity in the amygdala is dampened during emotion regulation, and such a modulation may change according to the specific strategy adopted (e.g., Goldin et al., 2008; Ochsner et al., 2012). Unfortunately, no modulation of activity in this structure emerged in our study. This result may be explained according to the finding of a recent meta-analytic study (Gentili et al., 2018) on the neural correlates of emotional stimuli processing in phobic patients vs healthy controls. The authors reported differences between the two groups only in the midcingulate cortex when exclusively whole-brain studies were selected. However, differences in several subcortical regions, among which the amygdala, emerged when also ROI-based studies were included. Finally, although we acknowledge that in meta-analytic studies a more stringent uncorrected cluster-forming threshold is commonly used (Müller et al., 2018), we also agree that this threshold is conventionally chosen despite "any other uncorrected voxel-wise thresholds would also be perfectly valid" (Eickhoff et al., 2012, pg. 2353-2354). We hope our preliminary, yet promising finding will expand neuroscientific investigations on emotion-focused strategies to offer a larger sample, and to consequently allow future metanalytic comparisons to apply more stringent parameters.

**Conclusions**

Reappraisal and acceptance are different effective processes to regulate emotions in responding to distressful events (Kohl et al., 2012; McRae, 2016; Aldao et al., 2010; Grecucci et al., 2020). In clinical psychology, the usefulness of such strategies is debated with different views concerning the usefulness of reappraisal to control emotion *versus* acceptance/non-controlling attitude toward emotions (Hofmann et al., 2009; Wolgast et al., 2011; Diedrich et al., 2016; Frederickson et al., 2018). With the present meta-analytic study, our aim was to contribute to this debate by shedding new light on the nature of common and specific patterns of brain activity associated with such processes. We believe that by comparing such opposite in nature strategies, the neural architecture of emotion regulation processes can be better outlined, by an exhaustive description of every facet of it.

**Funding**

First author's stipend was supported by a grant from the Italian Ministry of University and Research (Excellence Department Grant awarded to the Department of Psychology and Cognitive Science, University of Trento, Italy).

**Data and code availability statement**

The data analysed in the present meta-analytic study (i.e., the papers included in the meta-analyses) are cited in the references. The GingerALE output files, instead, are available online at the link https://osf.io/jf6z8/.

**Supplementary material**

Supplementary materials are available online at the link https://osf.io/jf6z8/.

Figures

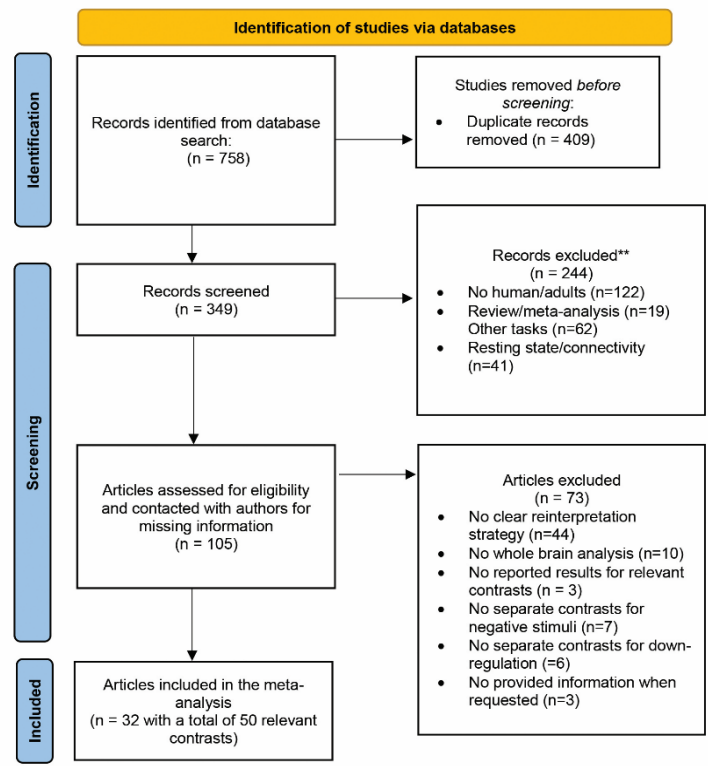

**Figure 1. Flowchart of the literature search and study selection process, based on PRISMA template (Page et al., 2021).**

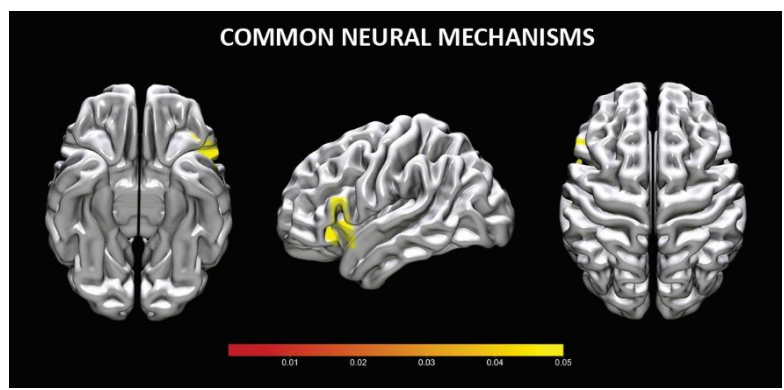

**Figure 2. Common neural mechanisms for reappraisal and acceptance. Coordinates are reported in MNI-space.**

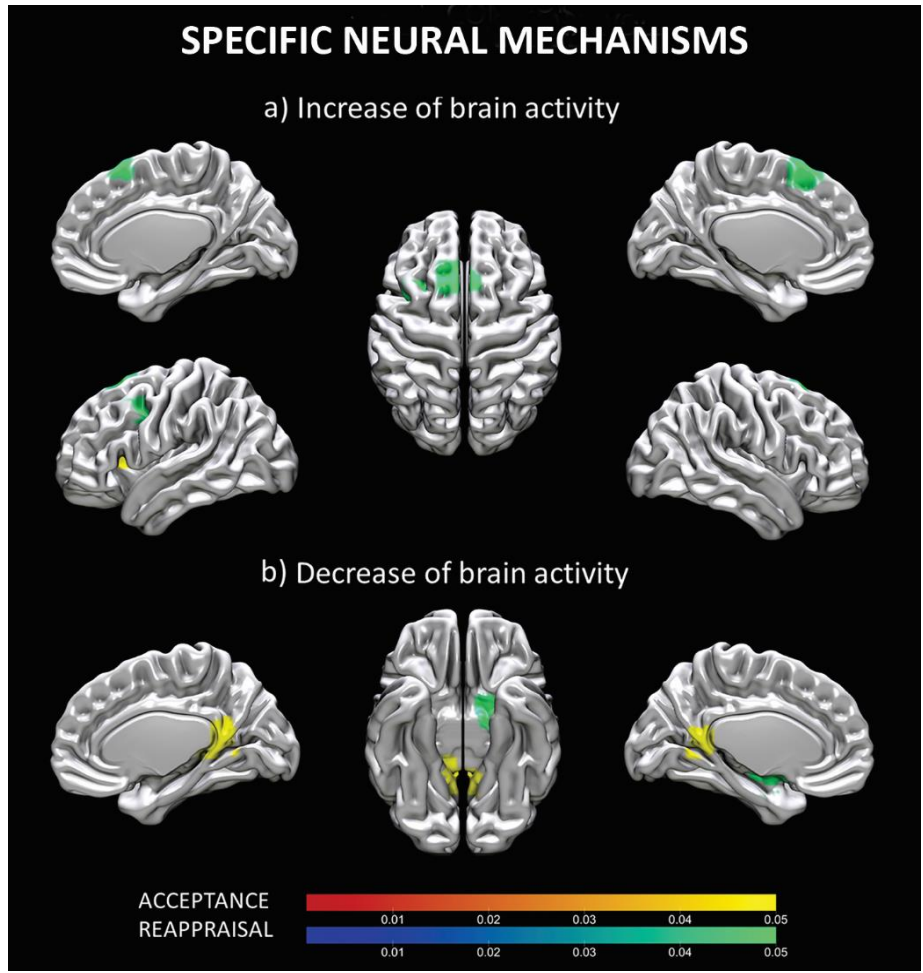

**Figure 3. Specific neural mechanisms for reappraisal and acceptance. Increased (a) and decreased (b) brain activity for regions specifically involved in acceptance (red-yellow scale) and reappraisal (Blue-green scale) strategies. Coordinates are reported in MNI-space.**

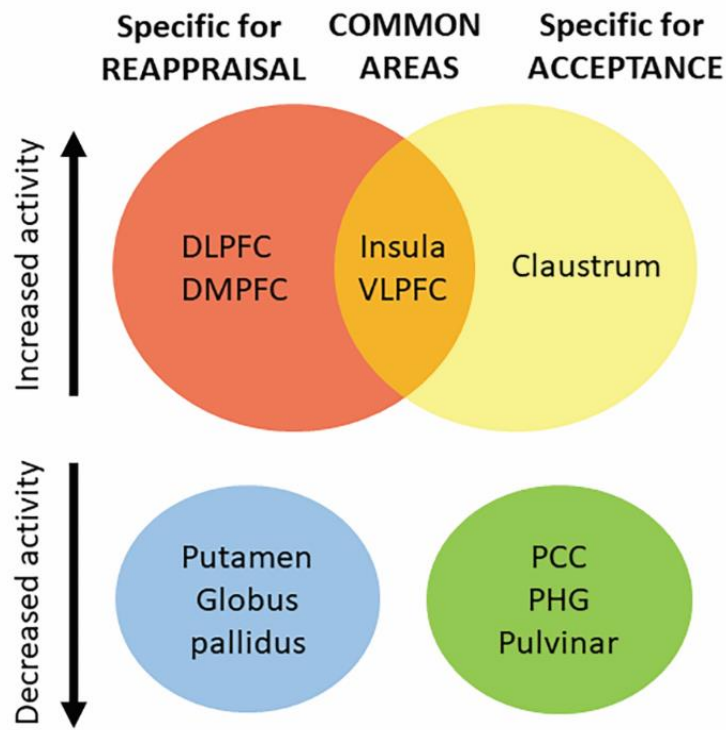

**Figure 4. Based on our results, we suggest that emotion regulation process relies on a set of common neural areas (central part of the figure), which coexists with strategy-specific mechanisms separating reappraisal-like strategies (on the left), from acceptance-like strategies (on the right). Top of the figure: areas showing increased brain activity; Bottom of the figure: areas showing decreased brain activity. VLPFC = Ventro-lateral Prefrontal Cortex; DLPFC= Dorso-lateral Prefrontal Cortex; DMPFC= dorso-medial Prefrontal Cortex; PCC = Posterior Cingulate Cortex; PHG = Parahippocampal gyrus; IFG = Inferior Frontal Gyrus.**